\documentclass[preprint,showpacs,showkeys,preprintnumbers,amsmath,amssymb,aps]{revtex4}

\usepackage{mathrsfs}
\usepackage{graphicx}
\usepackage{dcolumn}
\usepackage{bm}

\begin{document}

\preprint{APS/123-QED}

\title{Heun equation, Teukolsky equation, and type-D metrics}

\author{D.Batic}
\email{batic@itp.phys.ethz.ch}
\affiliation{ 
Institute for Theoretical Physics\\
Swiss Federal Institute of Technology\\
CH-8093 Z\"{u}rich, Switzerland}

\author{H.Schmid}
\affiliation{
UBH Software \& Engineering GmbH\\
D-92263 Amberg, Germany
}
\email{Harald.Schmid@UBH.de}

\date{\today} 

\begin{abstract}
Starting with the whole class of type-D vacuum backgrounds with cosmological constant we show that the separated Teukolsky equation for zero rest-mass fields with spin $s=\pm 2$ (gravitational waves), $s=\pm 1$ (electromagnetic waves) and $s=\pm 1/2$ (neutrinos) is an Heun equation in disguise.
\end{abstract}

\maketitle

\section{\label{sec:1}Introduction}
According to Ronveaux (1995) the Heun equation (HE) is the most general second order linear ODE of the form
\begin{equation}\label{Heun}
\frac{d^{2}y}{dz^2}+\left(\frac{1+2\alpha_{1}}{z}+\frac{1+2\alpha_{2}}{z-1}+\frac{1+2\alpha_{3}}{z-z_{S}}\right)\frac{dy}{dz}+\frac{\alpha\beta z-q_{A}}{z(z-1)(z-z_{S})}y=0
\end{equation}
where $\widehat{a}\in\mathbb{C}\backslash\{0,1\}$, $\alpha_{1}$, $\alpha_{2}$, $\alpha_{3}$, $\alpha$, $\beta$, $q_{A}$ are complex arbitrary parameters, and $0$, $1$, $\widehat{a}$, and $\infty$ are regular singularities with exponents $\{0,-2\alpha_{1}\}$, $\{0,-2\alpha_{2}\}$, $\{0,-2\alpha_{3}\}$, and $\{\alpha,\beta\}$, respectively. Equation \eqref{Heun} has been originally constructed by the German mathematician Karl Heun (1889) as a generalization of the hypergeometric equation (HYE). In order to see how HE degenerates to the HYE we can first multiply \eqref{Heun} by $z(z-1)(z-\widehat{a})$, then we set $\widehat{a}=1$, and $q_{A}=\alpha\beta$, and finally we take out a factor $(z-1)$, leaving the HYE in its standard form. Hence, we can always think to a HYE as a degenerated equation of Heun's type. To underline the importance of \eqref{Heun} we recall that it contains the generalized spheroidal equation (GSWE), the Coulomb spheroidal equation, Lam$\acute{\text{e}}$, Mathieau, and Ince equations as special cases. The fields of applications of the HE in physics are so large that it is not possible to describe them here in detail. However, a review of many general situations relevant to physics, chemistry, and engineering where the HE occurs can be found in Ronveaux (1995) (pp 341). Here, we will show which role plays the HE in quantum field theory in curved spacetimes.\\
To understand the motivation underlying the present work we give a short review on studies concerning exact solutions of the Teukolsky equation in some black hole geometries. In the 70's, and 80's we find a large number of publications regarding the angular equation obtained after separation of variables from the Teukolsky wave equation on Kerr manifolds. See, for instance, Press, and Teukolsky (1973), Breuer et al. (1977), Fackerell, and Crossman (1977), Leahy, and Unruh (1979), Chakrabarti (1984), and Seidel (1989). According to these references we will also name the solution of the angular equation as spin-weighted spheroidal function (SWSF). The radial equation has been investigated by Bardeen, and Press (1973), Page (1973), Lee (1976), Arenstorf et al. (1978). The common picture emerging from all previous studies is that the radial equation cannot be in general related to any known differential equation of mathematical physics. This view changed with the work of Blandin et al. (1983). They showed that the SWSF's may be obtained by means of an elementary transformation from Heun confluent functions. Three years later Leaver (1983) proved that the radial, and angular parts of the Teukolsky master equation (TME) in the Kerr geometry are generalized spheroidal wave equations. Finally, Suzuki et al. (1998) showed that the radial, and angular part equations arising from the TME in the Kerr-Newman-deSitter metric (KNdS) after separation of variables are Heun equations. It is interesting to observe that if we let the cosmological constant go to zero (i.e. the KNdS geometry goes over to the Kerr-Newman metric) their HEs become a confluent HE which, in turn, coincides with the GSWE given by Leaver in 1983. Hence, the following question arises quite naturally, namely: is it possible to reduce the TME in any physical relevant type D metric to a HE? To conclude this short review we cite what Wu, and Cai wrote in 2003: ''it is not clear until now whether the generalized Teukolsky equation in the general type D vacuum backgrounds with cosmological constant can be transformed into a Heun equation.''\\
Our paper is organized as follows: in Sec.~\ref{sec:2} we shortly present some results due to Kamran, and McLenaghan (1987) concerning the separation of the TME in any type-D background. In Sec.~\ref{sec:3}-\ref{sec:7} we show that the TME can be transformed in any physical relevant type D metric into a HE.    
\section{\label{sec:2}Background}
Let $\mathscr{D}_{0}$ denote the class of algebraically special Petrov type D vacuum metrics with cosmological constant. According to Thm. 2.1 in Kamran, and McLenaghan (1987) there exists a system of local coordinates $(u,v,w,x)$ in which such metrics can be written as
\begin{equation}\label{metric}
ds^{2}=2\left(\theta^{1}\theta^{2}-\theta^{3}\theta^{4}\right)
\end{equation}
with a symmetric null tetrad $(\theta^{1},\theta^{2},\theta^{3},\theta^{4})$ given by
\begin{eqnarray*}
\theta^{1}&=&\frac{\sqrt{Z(w,x)}}{\sqrt{2}T(w,x)}\left[\frac{fW(w)}{Z(w,x)}\left(\epsilon_{1}~du+m(x)~dv\right)+\frac{dw}{g^{2}W(w)}\right],\\
\theta^{2}&=&\frac{\sqrt{Z(w,x)}}{\sqrt{2}T(w,x)}\left[\frac{W(w)}{Z(w,x)}\left(\epsilon_{1}~du+m(x)~dv\right)-\frac{fdw}{g^{2}W(w)}\right],\\
\theta^{3}&=&\frac{\sqrt{Z(w,x)}}{\sqrt{2}T(w,x)}\left[\frac{X(x)}{Z(w,x)}\left(\epsilon_{2}~du+p(w)~dv\right)+i\frac{dx}{X(x)}\right]=\overline{\theta^{4}},\\
Z(w,x)&:=&\epsilon_{1}p(w)-\epsilon_{2}m(x),\quad g:=\sqrt{\frac{1+f^{2}}{2}}
\end{eqnarray*}
where all functions are real-valued and $\epsilon_{1}$, $\epsilon_{2}$, and $f$ are constants such that $\epsilon_{1}^{2}+\epsilon^{2}_{2}\neq 0$. Depending on whether $fW^{2}(w)$ is positive, negative or zero the metric \eqref{metric} possesses a two-parameter abelian group of isometries whose orbits are timelike, spacelike or null at a given point, respectively. By integration of the Einstein-Maxwell field equation it results that the general solution $\widetilde{\mathscr{A}}^{*}$ in the class $\mathscr{D}_{0}$ can be specified as follows (Thm. 2.2 ibid.)
\begin{eqnarray*}
\epsilon_{1}&=& b^{2}\cos{\gamma},\quad\epsilon_{2}=\sin{\gamma},\\
m(x)&=&-\left[c^{2}x^2+b^2k^2+\frac{\ell^{2}(1-b^2\cos^{2}{\gamma})}{\epsilon_{2}^{2}}\right]\epsilon_{2}-2c\ell x,\\
p(w)&=&+\left[b^{2}c^{2}w^2+\ell^2+\frac{k^{2}(b^2-\epsilon_{2}^{2})}{\cos^{2}{\gamma}}\right]\cos{\gamma}+2b^2ckw,\\
T(w,x)&=&a(cw\cos{\gamma}+k)(cx\sin{\gamma}+\ell)+1,\\
fW^{2}(w)&=&c^2b^4g_{4}w^4\cos^{2}{\gamma}+cf_{3}w^{3}\cos{\gamma}+f_{2}w^{2}+f_{1}w+f_{0},\\
X^{2}(x)&=&c^2g_{4}x^{4}\sin^{2}{\gamma}+c\kappa_{1}x^3\sin{\gamma}+\kappa_{2}x^2+g_{1}x+g_{0},\\
\kappa_{1}&:=&acf_{1}\cos{\gamma}-2akf_{2}+3ak^2f_{3}+4(\ell-ab^4k^3)g_{4},\\
\kappa_{2}&:=&3ac\ell f_{1}\cos{\gamma}-(1+6ak\ell)f_{2}+3k(1+3ak\ell)f_{3}+6[\ell^2-b^4k^2(1+2ak\ell)]g_{4}
\end{eqnarray*}
where $f_{0}$, $f_{1}$, $f_{2}$, $f_{3}$, $g_{0}$, $g_{1}$, $g_{4}$, $a$, $b$, $c$, $k$, $\ell$, and $\gamma$ are real parameters satisfying the relations
\begin{multline}\label{parametri0}
acg_{1}\sin{\gamma}-3a^2c\ell^2f_{1}\cos{\gamma}+2a\ell(1+3ak\ell)f_{2}-[1+3ak\ell(2+3ak\ell)]f_{3}\\
+4[b^4k-a\ell^3+3ab^4k^2\ell(1+ak\ell)]g_{4}=0,
\end{multline}
\begin{multline}\label{parametri1}
c^2(g_{0}\sin^{2}{\gamma}-b^{4}f_{0}\cos^{2}{\gamma})+c[(2a\ell^{3}+b^{4}k)f_{1}\cos{\gamma}-\ell g_{1}\sin{\gamma}]-(b^4k^2+\ell^2+4ak\ell^3)f_{2}\\
+(b^4k^3+3k\ell^2+6ak^2\ell^3)f_{3}+(3\ell^4-b^8k^4-6b^4k^2\ell^2-8ab^4k^3\ell^3)g_{4}=0,
\end{multline}
and are restricted to a range such that \eqref{metric} is non-singular with signature minus two. Moreover, the cosmological constant $\Lambda$ is expressed in terms of the above parameters by
\begin{equation}\label{cosmological}
\Lambda=-3[a^2c^2f_{0}\cos^{2}{\gamma}-a^2 ckf_{1}\cos{\gamma}+a^2k^2f_{2}-a^3k^3f_{3}+(1+a^2b^4k^4)g_{4}].
\end{equation}
Let $\mathscr{A}^{*}$ denote the subclass of solutions in $\widetilde{\mathscr{A}}^{*}$ obtained by setting $b=1$, $c=\sqrt{2}$, $k=\ell=0$ and $\gamma=\pi/4$. If in addition $a=f=1$, such solutions recover the vacuum case with cosmological constant of the seven-parameter family of Plebanski and Demianski (1976) containing the Kerr-Newman-de Sitter metric as a special case.\\
Let $\mathscr{B}^{0}_{-}$ be the subclass of solutions in $\widetilde{\mathscr{A}}^{*}$ such that $a=0$, $b=c=1$, $\ell=0$ and $\gamma=\pi/2$. If in addition $f=1$, $\mathscr{B}^{0}_{-}$ reduces to the vacuum case with cosmological constant of Carter's $\widetilde{B}_{-}$ (1968). By $\mathscr{B}^{0}_{+}$ we will denote the subclass of solutions in $\widetilde{\mathscr{A}}^{*}$ obtained by setting $a=0$, $b=c=1$, and $k=\gamma=0$. If we let $f=1$, such a solution becomes the vacuum case with cosmological constant of Carter's $\widetilde{B}_{+}$. The Carter's $\widetilde{B}_{\pm}$ describe all non-accelerating type D metrics in a coordinate system in which the components of the metric and the Maxwell field are rational functions. \\
Let $\mathscr{C}^{*}$ denote the subclass of solutions in $\widetilde{\mathscr{A}}^{*}$ obtained by setting $a=1$, $b=0$, $c=\sqrt{2}$, $k=\ell=0$ and $\gamma=\pi/4$. If in addition $f=1$, $\mathscr{C}^{*}$ reduces to the accelerating $\mathscr{C}$-metric of Levi-Civita (1918).\\
Finally, let $\mathscr{C}^{00}$ denote the subclass of solutions in $\widetilde{\mathscr{A}}^{*}$ obtained by setting $a=0$, $b=1$, $c=0$, $k=0$, $\ell=1$ and $\gamma=0$. If in addition $f=1$, $\mathscr{C}^{00}$ becomes the Robinson-Bertotti solution (1959).\\
Following Kamran, and McLenaghan (1987) the Teukolsky equation can be written in a compact form by introducing a spin parameter $s$ which can assume the values $\pm 2$, $\pm 1$, and $\pm 1/2$. For $s=2$, $1$ and $1/2$ we have
\begin{multline}\label{T1}
[(D-(2s-1)\epsilon+\overline{\epsilon}-2s\rho-\overline{\rho})(\Delta-2s\gamma+\mu)-(\delta+\overline{\pi}-\overline{\alpha}-(2s-1)\beta-2s\tau)(\overline{\delta}+\pi-2s\alpha)\\
-2(s-\frac{1}{2})(s-1)\Psi_{2}]\Phi_{s}=0,
\end{multline} 
and for $s=-2$, $-1$ and $-1/2$
\begin{multline}\label{T2}
[(\Delta-(2s+1)\gamma-\overline{\gamma}-2s\mu+\overline{\mu})(D-2s\epsilon-\rho)-(\overline{\delta}-\overline{\tau}+\overline{\beta}-(2s+1)\alpha-2s\pi)(\delta-\tau-2s\beta)\\
-2(s+\frac{1}{2})(s+1)\Psi_{2}]\Phi_{s}=0
\end{multline} 
where $\Psi_{2}$ is the non-zero Newman-Penrose component of the Weyl tensor and the $\Phi_{s}$'s are defined in terms of the field components as given in Table 1 (pp 286 ibid.). According to Thm. 3.1 (ibid.) for all solutions in the class $\mathscr{D}_{0}$, and for all $s=\pm 2,\pm 1,\pm 1/2$ equations \eqref{T1}, and \eqref{T2} possess a separable solution of the form
\[
\Phi_{s}=e^{i(ru+qv)}\frac{T^{|s|+1}}{Z^{|s|/2}}e^{i|s|B/2}\Theta_{s}(w,x)
\]
where $r$, $q$ are arbitrary real constants, $dB=Z^{-1}(\epsilon_{1}m^{'}dw+\epsilon_{2}~p^{'}dx)$, and $\Theta_{s}(w,x)=G_{s}(w)H_{s}(x)$. Moreover, \eqref{T1}, and \eqref{T2} separate into the pair of decoupled ODE's
\begin{eqnarray}
&&D_{ws}L_{ws}G_{s}(w)+f_{s}(w)G_{s}(w)=\lambda_{s}G_{s}(w),\label{radial}\\
&&D_{xs}L_{xs}H_{s}(x)-g_{s}(x)H_{s}(x)=\lambda_{s}H_{s}(x)\label{angular}
\end{eqnarray}
where $\lambda_{s}$ is a separation constant, the functions $f_{s}$, and $g_{s}$ are given in Table 1 (pp 290-291 ibid.), and 
\begin{eqnarray*}
D_{ws}&=&W\frac{d}{dw}+\frac{i}{1+f^2}\left[f\left(1+\epsilon(s)\right)-\frac{1}{f}\left(1-\epsilon(s)\right)\right]\frac{pr-\epsilon_{2}q}{W}+(1-|s|)\frac{dW}{dw},\\
D_{xs}&=&-iX\frac{d}{dx}+i\epsilon(s)\frac{\epsilon_{1}q-mr}{X}+i(|s|-1)\frac{dX}{dx},\\
L_{ws}&=&-fW\frac{d}{dw}+\frac{i}{1+f^2}\left[\left(1+\epsilon(s)\right)-f^2\left(1-\epsilon(s)\right)\right]\frac{pr-\epsilon_{2}q}{W}-|s|f\frac{dW}{dw},\\
L_{xs}&=&iX\frac{d}{dx}+i\epsilon(s)\frac{\epsilon_{1}q-mr}{X}+i|s|\frac{dX}{dx},
\end{eqnarray*}
where $\epsilon(s)$ is a sign function such that $\epsilon(s)=+1$ for $s>0$, and $\epsilon(s)=-1$ for $s<0$.
\section{\label{sec:3}The metric $\mathscr{A}^{*}$ : $b=1$,  $c=\sqrt{2}$,  $k=\ell=0$ , $\gamma=\pi/4$}
In this case we have
\begin{eqnarray*}
\epsilon_{1}&=&\epsilon_{2}=\frac{\sqrt{2}}{2},\quad m(x)=-\sqrt{2}~x^2,\quad p(w)=\sqrt{2}~w^2,\quad T(w,x)=awx+1,\\
fW^{2}(w)&=&g_{4}w^4+\sum_{n=0}^{3}f_{n}w^{n}=g_{4}\prod_{i=1}^{4}(w-w_{i}),\\
X^{2}(x)&=&g_{4}x^{4}+af_{1}x^3-f_{2}x^2+g_{1}x+g_{0}=g_{4}\prod_{i=1}^{4}(x-x_{i}).
\end{eqnarray*}
Moreover, for $i$,$j=1,\cdots,4$ let $w_{i}$ and $x_{i}$ denote the i-th root of the polynomial equations $fW^{2}(w)=0$, and $X^{2}(x)=0$, respectively. Throughout this section we shall assume that $w_{i}\neq w_{j}$, and $x_{i}\neq x_{j}$ for every $i\neq j$. Furthermore, in the present case \eqref{cosmological} reads
\[
g_{4}=-\left(a^2f_{0}+\frac{\Lambda}{3}\right).
\]
Making use of the expression for $\Psi_{2}$ given by (2.7g) in Kamran, and McLenaghan (1984), the functions $f_{s}(w)$, and $g_{s}(x)$ entering, respectively in \eqref{radial}, and \eqref{angular} are computed to be
\begin{eqnarray*}
&&f_{s}(w)=-(1-2|s|)\left[2g_{4}(1-|s|)w^2+\left(2i\sqrt{2}r\epsilon(s)+ag_{1}(1-|s|)\right)w\right],\\
&&g_{s}(x)=-(1-2|s|)\left[2g_{4}(1-|s|)x^2-\left(2\sqrt{2}r\epsilon(s)-af_{1}(1-|s|)\right)x\right].
\end{eqnarray*}
Hence, the equation for $G_{s}(w)$ becomes
\begin{equation}\label{uno_s}
\frac{d}{dw}\left(fW^2\frac{dG_{s}}{dw}\right)+\Gamma(w)\frac{dG_{s}}{dw}+Q_{s}(w)G_{s}=0
\end{equation}
with
\[
\Gamma(w):=i\sqrt{2}~\frac{f^2-1}{f^2+1}(2rw^2-q),
\]
\begin{multline*}
Q_{s}(w):=\lambda_{s}+2g_{4}(1-|s|)(1-2|s|)w^2+\left[2i\sqrt{2}r\left(-2s+\frac{f^2-1}{f^2+1}\right)+ag_{1}(1-|s|)(1-2|s|)\right]w+\\
f|s|\left(WW^{''}+(1-|s|){(W^{'})}^2\right)+i\sqrt{2}s(2rw^2-q)\frac{W^{'}}{W}+\frac{2f}{(1+f^2)^2}\frac{(2rw^2-q)^2}{W^2}.
\end{multline*}
Equation \eqref{uno_s} can be further simplified. To this aim let us make the substitution
\[
G_{s}(w)=e^{h(w)}\varphi_{s}(w).
\] 
If we require that $h^{'}=-\Gamma/(2fW^2)$ we obtain the following equation for $\varphi_{s}(w)$
\begin{equation}\label{due_s}
\frac{d}{dw}\left(fW^2\frac{d\varphi_{s}}{dw}\right)+R_{s}(w)~\varphi_{s}=0
\end{equation}
where
\begin{multline*}
R_{s}(w):=\lambda_{s}+2g_{4}(1-|s|)(1-2|s|)w^2-\mathcal{C}_{s}w+\\
f|s|\left(WW^{''}+(1-|s|){(W^{'})}^2\right)+i\sqrt{2}s(2rw^2-q)\frac{W^{'}}{W}+\frac{(2rw^2-q)^2}{2fW^2}
\end{multline*}
with 
\[
\mathcal{C}_{s}:=4i\sqrt{2}sr-ag_{1}(1-|s|)(1-2|s|).
\]
Let us introduce the following functions
\[
\sigma_{\pm q}(w):=2rw\pm q,\quad \mathfrak{f}_{s}(w):=2g_{4}(1-|s|)(1-2|s|)w^{2}-\mathcal{C}_{s}w+\lambda_{s},
\]
and let us define constants
\[
c_{i}^{-1}:=g_{4}\prod_{j=1 \atop j\neq i}^{3}(w_{i}-w_{j}),\quad i=1,2,3.
\]
By means of the homographic substitution
\begin{equation}\label{homog}
z=\frac{w-w_{1}}{w-w_{4}}~\frac{w_{2}-w_{4}}{w_{2}-w_{1}}
\end{equation}
mapping the points $w_{1},w_{2},w_{3},w_{4},\infty$ to $0,1,z_{S},\infty,z_{\infty}$ with
\begin{equation}\label{zI_zs}
z_{\infty}:=\frac{w_{2}-w_{4}}{w_{2}-w_{1}},\quad z_{S}:=\frac{w_{3}-w_{1}}{w_{3}-w_{4}}z_{\infty}
\end{equation}
equation \eqref{due_s} becomes
\begin{equation}\label{stella}
\frac{d^2\varphi_{s}}{dz^2}+P(z)\frac{d\varphi_{s}}{dz}+\widetilde{Q}_{s}(z)\varphi_{s}=0
\end{equation}
with
\begin{eqnarray*}
&&P(z)=\frac{1}{z}+\frac{1}{z-1}+\frac{1}{z-z_{S}}-\frac{2}{z-z_{\infty}},\\
&&\widetilde{Q}_{s}(z)=\frac{B_{1}}{z^2}+\frac{B_{2}}{(z-1)^2}+\frac{B_{3}}{(z-z_{S})^2}+\frac{2}{(z-z_{\infty})^2}+\frac{A_{1}}{z}+\frac{A_{2}}{z-1}+\frac{A_{3}}{z-z_{S}}+\frac{A_{\infty}}{z-z_{\infty}}
\end{eqnarray*}
where for $i=1,2,3$
\begin{eqnarray*}
&&A_{\infty}=\frac{1}{z_{\infty}(w_{4}-w_{1})}\left((2|s|^2-3|s|+2)\sum_{i=1}^{3}w_{i}+(1-|s|)(1-2|s|)\frac{3ag_{1}}{g_{4}}-(2-|s|)(1+2|s|)w_{4}\right),\\
&&B_{i}=-\left(\frac{s}{2}-i\frac{c_{i}\sigma_{-q}(w_{i}^2)}{\sqrt{2}(w_{i}-w_{4})}\right)^2,\\
&&A_{1}=-\frac{c_{1}}{z_{\infty}}~\mathfrak{g}_{s}(w_{1}),\quad A_{2}=c_{2}\frac{w_{2}-w_{1}}{w_{4}-w_{1}}~\mathfrak{g}_{s}(w_{2}),\quad A_{3}=c_{3}\frac{w_{3}-w_{1}}{z_{S}(w_{4}-w_{1})}~\mathfrak{g}_{s}(w_{3}),
\end{eqnarray*}
and
\begin{multline*}
\mathfrak{g}_{s}(w_{i})=\frac{g_{4}}{2}|s|p(w_{i})-\mathfrak{f}_{s}(w_{i})+i\frac{\sqrt{2}s}{w_{i}-w_{4}}~\sigma_{+q}\left(w_{i}(w_{4}-2w_{i})\right)+\\
\frac{g_{4}c_{i}^{2}}{w_{i}-w_{4}}\sigma_{-q}(w_{i}^2)\left[\sigma_{+q}(w_{i}^2)\sum_{j=1 \atop j\neq i}^{3}w_{j}-2w_{i}\sigma_{+q}\left(\prod_{j=1 \atop j\neq i}^{3}w_{j}~\right)\right],
\end{multline*}
\begin{equation}\label{polp}
p(w_{i}):=2(2|s|-3)w_{i}^2+(4-3|s|)w_{i}\sum_{j=1 \atop j\neq i}^{3}w_{j}+(|s|-2)w_{4}\left(\sum_{j=1 \atop j\neq i}^{3}w_{j}-2w_{i}\right)+2(|s|-1)\prod_{j=1 \atop j\neq i}^{3}w_{j}.
\end{equation}
If we make the F-homotopic transformation
\[
\varphi_{s}(z)=z^{\alpha_{1}}(z-1)^{\alpha_{2}}(z-z_{S})^{\alpha_{3}}(z-z_{\infty})\widetilde{\varphi}_{s}(z)
\]
and require that $\alpha_{i}^{2}=-B_{i}$, then \eqref{stella} becomes
\begin{equation}\label{stellastella}
\frac{d^2\widetilde{\varphi}_{s}}{dz^2}+\widehat{P}_{s}(z)\frac{d\widetilde{\varphi}_{s}}{dz}+\widehat{Q}_{s}(z)\widetilde{\varphi}_{s}=0
\end{equation}
with
\[
\widehat{P}_{s}(z)=\frac{1+2\alpha_{1}}{z}+\frac{1+2\alpha_{2}}{z-1}+\frac{1+2\alpha_{3}}{z-z_{S}},\quad
\widehat{Q}_{s}(z)=\frac{\widehat{A}_{1}}{z}+\frac{\widehat{A}_{2}}{z-1}+\frac{\widehat{A}_{3}}{z-z_{S}}+\frac{\widehat{A}_{\infty}}{z-z_{\infty}}.
\]
Now, a direct computation gives
\begin{eqnarray}
&&\widehat{A}_{\infty}=\frac{(1-|s|)(1-2|s|)}{z_{\infty}(w_{4}-w_{1})}\left(\sum_{i=1}^{4}w_{i}+\frac{ag_{1}}{g_{4}}\right), \label{c1}\\
&&\sum_{i=1}^{3}\widehat{A}_{i}=-\frac{(1-|s|)(1-2|s|)}{z_{\infty}(w_{4}-w_{1})}\left(\sum_{i=1}^{4}w_{i}+\frac{ag_{1}}{g_{4}}\right). \label{c2}
\end{eqnarray}
Taking into account that for the metric $\mathscr{A}^{*}$ equation \eqref{parametri0} reduces to $ag_{1}-f_{3}=0$, and replacing $f_{3}$ by $ag_{1}$ in the equation $fW^{2}=0$ it can be checked that the following relation holds
\[
\sum_{i=1}^{4}w_{i}=-\frac{ag_{1}}{g_{4}}.
\]
Hence, the coefficients \eqref{c1}, and \eqref{c2} are zero, and \eqref{stellastella} reduces to an Heun equation with terms $\alpha\beta$ and accessory parameter $q_{A}$ (according to the notation in \eqref{Heun}) given by
\begin{equation}\label{parametri}
\alpha\beta=-(z_{S}+1)\widehat{A}_{1}-z_{S}\widehat{A}_{2}-\widehat{A}_{3}=z_{S}\widehat{A}_{3}+\widehat{A}_{2},\quad q_{A}=-z_{S}\widehat{A}_{1}
\end{equation}
where
\begin{eqnarray}
\widehat{A}_{1}&=&A_{1}-\alpha_{1}-\alpha_{2}(1+2\alpha_{1})-\frac{\alpha_{1}+\alpha_{3}(1+2\alpha_{1})}{z_{S}}-\frac{1}{z_{\infty}},\label{I}\\
\widehat{A}_{2}&=&A_{2}+\alpha_{1}+\alpha_{2}(1+2\alpha_{1})-\frac{\alpha_{2}+\alpha_{3}(1+2\alpha_{2})}{z_{S}-1}-\frac{1}{z_{\infty}-1},\label{II}\\
\widehat{A}_{3}&=&A_{3}+\frac{\alpha_{2}+\alpha_{3}(1+2\alpha_{2})}{z_{S}-1}+\frac{\alpha_{1}+\alpha_{3}(1+2\alpha_{1})}{z_{S}}-\frac{1}{z_{\infty}-z_{s}}\label{III}.
\end{eqnarray}
Concerning \eqref{angular} we find the following equation for $H_{s}(x)$, namely
\begin{multline}\label{unouno}
\frac{d}{dx}\left(X^2\frac{dH_{s}}{dx}\right)+\left[-\lambda_{s}+2g_{4}(1-|s|)(1-2|s|)x^2+\widehat{\mathcal{C}}_{s}x+\right.\\
\left.|s|\left(XX^{''}-(|s|-1){X^{'}}^2\right)-\sqrt{2}s(2rx^2+q)\frac{X^{'}}{X}-\frac{(2rx^2+q)^2}{2X^2}\right]H_{s}=0
\end{multline}
with
\[
\widehat{\mathcal{C}}_{s}:=4\sqrt{2}rs+(1-|s|)(1-2|s|)af_{1}.
\]
Let us introduce the following notation
\begin{eqnarray*}
&&\widehat{\mathfrak{f}}_{s}(x)=2g_{4}(1-|s|)(1-2|s|)x^{2}+\widehat{\mathcal{C}}_{s}x-\lambda_{s},\\
&&\widehat{c}_{i}~^{-1}=g_{4}\prod_{j=1 \atop j\neq i}^{3}(x_{i}-x_{j}),\quad i=1,2,3.
\end{eqnarray*}
By means of the homographic substitution
\begin{equation}\label{homog2}
z=\frac{x-x_{1}}{x-x_{4}}~\frac{x_{2}-x_{4}}{x_{2}-x_{1}}
\end{equation}
mapping the points $x_{1},x_{2},x_{3},x_{4},\infty$ to $0,1,x_{S},\infty,x_{\infty}$ with
\begin{equation}\label{xI_xs}
x_{\infty}=\frac{x_{2}-x_{4}}{x_{2}-x_{1}},\quad x_{S}=\frac{x_{3}-x_{1}}{x_{3}-x_{4}}x_{\infty}
\end{equation}
our equation \eqref{unouno} becomes
\begin{equation}\label{stellang}
\frac{d^2H_{s}}{dz^2}+\mathfrak{P}_{s}(z)\frac{dH_{s}}{dz}+\mathfrak{Q}_{s}(z)H_{s}=0
\end{equation}
with
\begin{eqnarray*}
&&\mathfrak{P}_{s}(z)=\frac{1}{z}+\frac{1}{z-1}+\frac{1}{z-x_{S}}-\frac{2}{z-x_{\infty}},\\
&&\mathfrak{Q}_{s}(z)=\frac{\widetilde{B}_{1}}{z^2}+\frac{\widetilde{B}_{2}}{(z-1)^2}+\frac{\widetilde{B}_{3}}{(z-x_{S})^2}+\frac{2}{(z-x_{\infty})^2}+\frac{\widetilde{A}_{1}}{z}+\frac{\widetilde{A}_{2}}{z-1}+\frac{\widetilde{A}_{3}}{z-x_{S}}+\frac{\widetilde{A}_{\infty}}{z-x_{\infty}}
\end{eqnarray*}
where for $i=1,2,3$
\begin{eqnarray*}
&&\widetilde{A}_{\infty}=\frac{1}{x_{\infty}(x_{4}-x_{1})}\left((2|s|^2-3|s|+2)\sum_{i=1}^{3}x_{i}+(1-|s|)(1-2|s|)\frac{3ag_{1}}{g_{4}}-(2-|s|)(1+2|s|)x_{4}\right),\\
&&\widetilde{B}_{i}=-\left(\frac{s}{2}+\frac{\widehat{c}_{i}~\sigma_{+q}(x_{i}^2)}{\sqrt{2}(x_{i}-x_{4})}\right)^2,\\
&&\widetilde{A}_{1}=-\frac{\widehat{c}_{1}}{x_{\infty}}~\widehat{\mathfrak{g}}_{s}(x_{1}),\quad \widetilde{A}_{2}=\widehat{c}_{2}\frac{x_{2}-x_{1}}{x_{4}-x_{1}}~\widehat{\mathfrak{g}}_{s}(x_{2}),\quad \widetilde{A}_{3}=\widehat{c}_{3}\frac{x_{3}-x_{1}}{x_{S}(x_{4}-x_{1})}~\widehat{\mathfrak{g}}_{s}(x_{3}),
\end{eqnarray*}
and
\begin{multline*}
\widehat{\mathfrak{g}}_{s}(x_{i})=\frac{g_{4}}{2}|s|\widetilde{p}(x_{i})-\widehat{\mathfrak{f}}_{s}(x_{i})-\frac{\sqrt{2}s}{x_{i}-x_{4}}~\sigma_{-q}\left(x_{i}(x_{4}-2x_{i})\right)-\\
\frac{g_{4}\widehat{c}_{i}^{2}}{x_{i}-x_{4}}\sigma_{+q}(x_{i}^2)\left[\sigma_{-q}(x_{i}^2)\sum_{j=1 \atop j\neq i}^{3}x_{j}-2x_{i}\sigma_{-q}\left(\prod_{j=1 \atop j\neq i}^{3}x_{j}~\right)\right],
\end{multline*}
\begin{equation}\label{tilde_p}
\widetilde{p}(x_{i}):=2(2|s|-3)x_{i}^2+(4-3|s|)x_{i}\sum_{j=1 \atop j\neq i}^{3}x_{j}+(|s|-2)x_{4}\left(\sum_{j=1 \atop j\neq i}^{3}x_{j}-2x_{i}\right)+2(|s|-1)\prod_{j=1 \atop j\neq i}^{3}x_{j}.
\end{equation}
If we make the F-homotopic transformation
\[
H_{s}(z)=z^{\widehat{\alpha}_{1}}(z-1)^{\widehat{\alpha}_{2}}(z-x_{S})^{\widehat{\alpha}_{3}}(z-x_{\infty})\widetilde{H}_{s}(z)
\]
together with the requirement that $\widehat{\alpha}_{i}^{2}=-\widetilde{B}_{i}$, then \eqref{stellang} becomes
\begin{equation}\label{stellastellang}
\frac{d^2\widetilde{H}_{s}}{dz^2}+\widehat{\mathfrak{P}}_{s}(z)\frac{d\widetilde{H}_{s}}{dz}+\widehat{\mathfrak{Q}}_{s}(z)\widetilde{H}_{s}=0
\end{equation}
with
\[
\widehat{\mathfrak{P}}_{s}(z)=\frac{1+2\widehat{\alpha}_{1}}{z}+\frac{1+2\widehat{\alpha}_{2}}{z-1}+\frac{1+2\widehat{\alpha}_{3}}{z-x_{S}},\quad
\widehat{\mathfrak{Q}}_{s}(z)=\frac{\mathfrak{A}_{1}}{z}+\frac{\mathfrak{A}_{2}}{z-1}+\frac{\mathfrak{A}_{3}}{z-x_{S}}+\frac{\mathfrak{A}_{\infty}}{z-x_{\infty}}.
\]
Now, a direct computation gives
\begin{eqnarray}
&&\mathfrak{A}_{\infty}=\frac{(1-|s|)(1-2|s|)}{x_{\infty}(x_{4}-x_{1})}\left(\sum_{i=1}^{4}x_{i}+\frac{af_{1}}{g_{4}}\right),\label{c3}\\
&&\sum_{i=1}^{3}\mathfrak{A}_{i}=-\frac{(1-|s|)(1-2|s|)}{x_{\infty}(x_{4}-x_{1})}\left(\sum_{i=1}^{4}x_{i}+\frac{af_{1}}{g_{4}}\right)\label{c4}.
\end{eqnarray}
Going back to the polynomial equation $X^{2}(x)=0$ it can be verified that
\[
\sum_{i=1}^{4}x_{i}=-\frac{af_{1}}{g_{4}}.
\]
This implies that \eqref{c3}, and \eqref{c4} are zero, and \eqref{stellastellang} becomes an Heun equation with
\begin{equation}\label{alphabetaq}
\alpha\beta=x_{S}\mathfrak{A}_{3}+\mathfrak{A}_{2},\quad q_{A}=-x_{S}\mathfrak{A}_{1}
\end{equation}
where the $\mathfrak{A}_{i}$'s can be directly obtained from the formulae \eqref{I}-\eqref{III} for the coefficients $\widehat{A}_{i}$ by means of the formal substitutions $\widehat{A}_{i}\longrightarrow\mathfrak{A}_{i}$, $A_{i}\longrightarrow \widetilde{A}_{i}$, $\alpha_{i}\longrightarrow \widehat{\alpha}_{i}$, $z_{S}\longrightarrow x_{S}$ and $z_{\infty}\longrightarrow x_{\infty}$.

\section{\label{sec:4}The metric $\mathscr{B}^{0}_{-}$ : $b=c=1$,  $a=\ell=0$ , $\gamma=\pi/2$}
In the present case we have
\begin{eqnarray*}
\epsilon_{1}&=&0,\quad\epsilon_{2}=1,\quad m(x)=-(x^2+k^2),\quad p(w)=2kw,\quad T(w,x)=1,\\
fW^{2}(w)&=&f_{2}w^2+f_{1}w+f_{0}=f_{2}(w-w_{1})(w-w_{2}),\\
X^{2}(x)&=&g_{4}x^{4}+\kappa_{2}x^2+g_{1}x+g_{0}=g_{4}\prod_{i=1}^{4}(x-x_{i}),\quad\kappa_{2}=-f_{2}+3kf_{3}-6k^2g_{4}.
\end{eqnarray*}
Moreover, for $i$,$j=1,\cdots,4$ let $w_{1}$, $w_{2}$, and $x_{i}$ denote the roots of the polynomial equations $fW^{2}(w)=0$, and $X^{2}(x)=0$, respectively. Throughout this section we shall assume that $w_{1}\neq w_{2}$, and $x_{i}\neq x_{j}$ for every $i\neq j$. Finally, let us recall that in the present situation \eqref{parametri0}, \eqref{parametri1} and \eqref{cosmological} read, respectively
\begin{equation}\label{fur}
4kg_{4}-f_{3}=0,\quad g_{0}-k^{2}(k^2g_{4}-kf_{3}+f_{2})=0,\quad
\Lambda=-3g_{4}.
\end{equation}
Since the term $x^3$ is not present in the expression for $X^2$ it can be easily checked that the roots of the polynomial equation $X^2(x)=0$ satisfy the condition $x_1+x_2+x_3+x_4=0$. Taking into account the expression for $\Psi_2$ given by (2.8g) in Kamran, and McLenaghan (1984), the functions $f_s(w)$ and $g_{s}(x)$ entering, respectively in \eqref{radial}, and \eqref{angular} are computed to be 
\begin{eqnarray*}
&&f_{s}(w)=0,\\
&&g_{s}(x)=-2ir\epsilon(s)(1-2|s|)(k+ix)+(1-|s|)(1-2|s|)\left[i\frac{({X^2}(x))^{'}}{k-ix}-\frac{6X^2(x)+C_{\Lambda}(k+ix)}{3(k-ix)^2}\right]
\end{eqnarray*}
with $C_{\Lambda}:=-2k(3f_2+4\Lambda k^2)+3ig_1$. The equation for $G_{s}(w)$ becomes
\begin{equation}\label{unoB0m}
\frac{d}{dw}\left(fW^2\frac{dG_{s}}{dw}\right)+\widehat{\Gamma}(w)\frac{dG_{s}}{dw}+Q_{s}(w)G_{s}=0
\end{equation}
with
\[
\widehat{\Gamma}(w):=2i\frac{f^2-1}{f^2+1}(2rkw-q),
\]
\begin{multline*}
Q_{s}(w):=\lambda_{s}-\frac{2irk}{1+f^2}\left(1+\epsilon(s)-f^2(1-\epsilon(s))\right)+f|s|\left(WW^{''}+(1-|s|){W^{'}}^2\right)+\\
2is(2rkw-q)\frac{W^{'}}{W}+\frac{4f}{(1+f^2)^2}\frac{(2rkw-q)^2}{W^2}.
\end{multline*}
In order to simplify \eqref{unoB0m} we make the transformation
\[
G_{s}(w)=e^{\widehat{h}(w)}\varphi_{s}(w)
\] 
requiring that $\widehat{h}^{'}=-\widehat{\Gamma}/(2fW^2)$. Hence, we end up with the following equation for $\varphi_{s}(w)$
\begin{equation}\label{dueB0m}
\frac{d}{dw}\left(fW^2\frac{d\varphi_{s}}{dw}\right)+R_{s}(w)\varphi_{s}=0
\end{equation}
where
\[
R_{s}(w):=\lambda_{s}-2irk\epsilon(s)+f|s|\left(WW^{''}+(1-|s|){W^{'}}^2\right)+2is(2rkw-q)\frac{W^{'}}{W}+\frac{(2rkw-q)^2}{fW^2}.
\]
By means of the transformation
\begin{equation}\label{homogB0m}
z=\frac{w-w_{1}}{w_{2}-w_{1}}
\end{equation}
mapping the points $w_{1},w_{2},\infty$ to $0,1,\infty$ our equation \eqref{dueB0m} becomes
\begin{equation}\label{stellaB0m}
\frac{d^2\varphi_{s}}{dz^2}+P_{s}(z)\frac{d\varphi_{s}}{dz}+\widetilde{Q}_{s}(z)\varphi_{s}=0
\end{equation}
with
\[
P_{s}(z)=\frac{1}{z}+\frac{1}{z-1},\quad
\widetilde{Q}_{s}(z)=\frac{B_{1}}{z^2}+\frac{B_{2}}{(z-1)^2}+\frac{A_{1}}{z}+\frac{A_{2}}{z-1}
\]
where
\begin{eqnarray*}
&&B_{1}=-\left(\frac{s}{2}+i\frac{2rkw_{1}-q}{f_{2}(w_{2}-w_{1})}\right)^2,\quad B_{2}=-\left(\frac{s}{2}-i\frac{2rkw_{2}-q}{f_{2}(w_{2}-w_{1})}\right)^2,\\
&&A_{1}=-\frac{\lambda_{s}+2irk(\epsilon(s)-s)}{f_{2}}+\frac{|s|(|s|-2)}{2}+\frac{2}{f_{2}^2(w_2-w_1)^2}\prod_{i=1}^{2}(2rkw_{i}-q),\quad A_{2}=-A_{1}.
\end{eqnarray*}
If we make the F-homotopic transformation
\[
\varphi_{s}(z)=z^{\alpha_{1}}(z-1)^{\alpha_{2}}\widetilde{\varphi}_{s}(z)
\]
with the requirement that ${\alpha_{i}}^{2}=-B_{i}$ then \eqref{stellaB0m} reads
\begin{equation}\label{stellastellaB0m}
\widetilde{\varphi}_{s}^{''}(z)+\widehat{P}_{s}(z)\widetilde{\varphi}_{s}^{'}(z)+\widehat{Q}_{s}(z)\widetilde{\varphi}_{s}(z)=0
\end{equation}
with
\[
\widehat{P}_{s}(z)=\frac{1+2\alpha_{1}}{z}+\frac{1+2\alpha_{2}}{z-1},\quad
\widehat{Q}_{s}(z)=\frac{(A_{1}+A_{2})z+\alpha_{2}+\alpha_{1}(1+2\alpha_{2})-A_{1}}{z(z-1)}.
\]
Since $A_{2}=-A_{1}$ it follows that \eqref{stellastellaB0m} is an HYE. Notice that an HYE can be always thought as a degenerate Heun equation since it can be recovered from \eqref{Heun} by setting for instance $z_S=1$ and $q_{A}=\alpha\beta$ or $z_S=q=0$ or $1+2\alpha_3=0$ and $q=z_{S}~\alpha\beta$. Concerning the ODE \eqref{angular} for  $H_{s}(x)$ we find 
\begin{multline}\label{unouno_x}
\frac{d}{dx}\left(X^2\frac{dH_{s}}{dx}\right)+\left[-\left(\lambda_{s}-2irk\epsilon(s)(1-2|s|)\right)+4rsx+|s|\left(XX^{''}+(1-|s|){X^{'}}^2\right)-\right.\\
\left.2rs(x^2+k^2)\frac{X^{'}}{X}-r^2\frac{(x^2+k^2)^2}{X^2}+\right.\\
\left.(1-|s|)(1-2|s|)\left(\frac{(X^2)^{'}}{x+ik}-\frac{6X^2+iC_{\Lambda}(x-ik)}{3(x+ik)^2}\right)\right]H_{s}=0.
\end{multline}
Let us introduce the following function
\[
\sigma_{\pm}(x):=x\pm k^2.
\]
Moreover, let $\widehat{c}_{i}$ be defined as in the previous section. By means of the homographic substitution \eqref{homog2} equation \eqref{unouno_x} becomes
\begin{equation}\label{stellang_x}
\frac{d^2H_{s}}{dz^2}+\mathfrak{P}_{s}(z)\frac{dH_{s}}{dz}+\mathfrak{Q}_{s}(z)H_{s}=0
\end{equation}
with
\[
\mathfrak{P}_{s}(z)=\frac{1}{z}+\frac{1}{z-1}+\frac{1}{z-x_{S}}-\frac{2}{z-x_{\infty}},
\]
\begin{multline*}
\mathfrak{Q}_{s}(z)=\frac{\widetilde{B}_{1}}{z^2}+\frac{\widetilde{B}_{2}}{(z-1)^2}+\frac{\widetilde{B}_{3}}{(z-x_{S})^2}+\frac{2}{(z-x_{\infty})^2}+\frac{\widetilde{B}_{k}}{(z-x_{k})^2}+\\
\frac{\widetilde{A}_{1}}{z}+\frac{\widetilde{A}_{2}}{z-1}+\frac{\widetilde{A}_{3}}{z-x_{S}}+\frac{\widetilde{A}_{\infty}}{z-x_{\infty}}+\frac{\widetilde{A}_{k}}{z-x_{k}},\quad x_{k}:=x_{\infty}\frac{x_{1}+ik}{x_{4}+ik}
\end{multline*}
where for $i=1,2,3$
\begin{eqnarray*}
&&\widetilde{A}_{\infty}=\frac{1}{x_{\infty}(x_{4}-x_{1})}\left(\sum_{i=1}^{3}x_{i}-3x_{4}\right),\\
&&\widetilde{B}_{i}=-\left(\frac{s}{2}+\frac{r\widehat{c}_{i}~\sigma_{k}^{+}(x_{i}^2)}{x_{i}-x_{4}}\right)^2,\quad \widetilde{B}_{k}=-2(1-|s|)(1-2|s|)\left(1+\frac{kC_{\Lambda}}{3g_{4}}\prod_{j=1}^{4}(x_j+ik)^{-1}\right),\\
&&\widetilde{A}_{1}=-\frac{\widehat{c}_{1}}{x_\infty}~\widehat{\mathfrak{g}}_s(x_{1}),\quad \widetilde{A}_{2}=\widehat{c}_{2}\frac{x_{2}-x_{1}}{x_{4}-x_{1}}~\widehat{\mathfrak{g}}_s(x_{2}),\quad \widetilde{A}_{3}=\frac{\widehat{c}_{3}(x_{3}-x_{1})}{x_{S}(x_{4}-x_{1})}~\widehat{\mathfrak{g}}_s(x_{3}),\\
&&\widetilde{A}_{k}=\frac{(1-|s|)(1-2|s|)(x_4+ik)}{x_{\infty}(x_1-x_4)}\left(\tau (k)+i\frac{C_{\Lambda}}{3g_{4}}p(k)\prod_{j=1}^{4}(x_{j}+ik)^{-1}\right)\prod_{i=1}^{3}(x_{i}+ik)^{-1},
\end{eqnarray*}
with
\begin{multline*}
\widehat{\mathfrak{g}}_s(x_{i})=\lambda_{s}-2irk\epsilon(s)(1-2|s|)+\frac{g_4}{2}|s|\widetilde{p}(x_{i})+\frac{2rs}{x_{i}-x_{4}}\sigma_{+}(x_{i}x_4)-\frac{2g_{4}\widehat{c}_{i}^{2}r^2}{x_{i}-x_{4}}\sigma_{+}(x_{i}^2)\cdot\\
\cdot\left[\sigma_{-}(x_{i}^2)\sum_{j=1 \atop j\neq i}^{3}x_{j}-2x_{i}\sigma_{-}\left(\prod_{j=1 \atop j\neq i}^{3}x_{j}~\right)\right]-\frac{(1-|s|)(1-2|s|)}{x_{i}+ik}\left(\frac{x_i-x_4}{\widehat{c}_{i}}-\frac{i}{3}C_{\Lambda}\frac{x_i-ik}{x_i+ik}\right),
\end{multline*}
\begin{multline*}
\tau(k):=\left(3x_{4}-\sum_{i=1}^{3}x_{i}\right)k^2+2i(x_{1}x_{2}+x_{1}x_{3}-x_{2}x_{4}-x_{3}x_{4}-x_{1}x_{4}+x_{2}x_{3})k+\\
3x_{1}x_{2}x_{3}-x_{2}x_{3}x_{4}-x_{1}x_{2}x_{4}-x_{1}x_{3}x_{4},
\end{multline*}
\begin{multline*}
p(k):=3k^4-\left(\sum_{i=1}^{3}x_{i}+5x_4\right)k^3+\left(x_{1}x_{2}+x_{2}x_{3}+x_{1}x_{3}-3x_{4}\sum_{i=1}^{3}x_{i}\right)k^2+\\
i(x_1x_2x_4+x_1x_3x_4+x_2x_3x_4-3x_1x_2x_3)k-x_1x_2x_3x_4,
\end{multline*}
and $\widetilde{p}(x_{i})$ given by \eqref{tilde_p}. Inserting the roots of the polynomial equation $X^2(x)=0$ into the expressions for $\widetilde{B}_k$, and $\widetilde{A}_k$ we checked with the help of Maple 9.5 that $\widetilde{B}_k=\widetilde{A}_k=0$. Let us work with the F-homotopic transformation
\[
H_{s}(z)=z^{\widehat{\alpha}_{1}}(z-1)^{\widehat{\alpha}_{2}}(z-x_{S})^{\widehat{\alpha}_{3}}(z-x_{\infty})\widetilde{H}_{s}(z).
\]
Moreover, we require that ${\widehat{\alpha}_{i}}^{2}=-\widetilde{B}_{i}$. Then \eqref{stellang_x} becomes 
\begin{equation}\label{stellastellang_x}
\widetilde{H}_{s}^{''}(z)+\widehat{\mathfrak{P}}_{s}(z)\widetilde{H}_{s}^{'}(z)+\widehat{\mathfrak{Q}}_{s}(z)\widetilde{H}_{s}(z)=0
\end{equation}
with
\[
\widehat{\mathfrak{P}}_{s}(z)=\frac{1+2\widehat{\alpha}_{1}}{z}+\frac{1+2\widehat{\alpha}_{2}}{z-1}+\frac{1+2\widehat{\alpha}_{3}}{z-x_{s}},\quad
\widehat{\mathfrak{Q}}_{s}(z)=\frac{\mathfrak{A}_{1}}{z}+\frac{\mathfrak{A}_{2}}{z-1}+\frac{\mathfrak{A}_{3}}{z-x_{s}}.
\]
Taking into account that the $\mathfrak{A}_{i}$'s can be directly obtained from the formulae \eqref{I}-\eqref{III} for the coefficients $\widehat{A}_{i}$ by means of the formal substitutions $\widehat{A}_{i}\longrightarrow\mathfrak{A}_{i}$, $A_{i}\longrightarrow \widetilde{A}_{i}$, $\alpha_{i}\longrightarrow \widehat{\alpha}_{i}$, $z_{S}\longrightarrow x_{S}$ and $z_{\infty}\longrightarrow x_{\infty}$, and that  $\widetilde{A}_{k}=0$ implies that
\[
\tau (k)+i\frac{C_{\Lambda}}{3g_{4}}p(k)\prod_{j=1}^{4}(x_{j}+ik)^{-1}=0,
\]
we obtain
\begin{eqnarray*}
\sum_{i=1}^{3}\mathfrak{A}_{i}&=&\sum_{i=1}^{3}\widetilde{A}_{i}-\left(\frac{1}{x_{\infty}}+\frac{1}{x_{\infty}-x_{S}}+\frac{1}{x_{\infty}-1}\right)\\
&=&\frac{(1-|s|)(1-2|s|)(x_1-x_2)}{(x_4-x_2)(x_1-x_4)}\left[3x_4-\sum_{i=1}^{3}x_i+\right.\\
&&\left.\prod_{i=1}^{3}(x_i+ik)^{-1}\left(T(k)-i\frac{C_{\Lambda}}{3g_{4}}p(k)(x_4+ik)\prod_{j=1}^{4}(x_{j}+ik)^{-1}\right)\right]\\
&=&\frac{(1-|s|)(1-2|s|)(x_1-x_2)}{(x_4-x_2)(x_1-x_4)}M(k)
\end{eqnarray*}
with
\begin{equation}\label{emme}
M(k):=3x_4-\sum_{i=1}^{3}x_i+\left(T(k)+(x_4+ik)\tau(k)\right)\prod_{i=1}^{3}(x_i+ik)^{-1},
\end{equation}
\begin{multline*}
T(k):=\left(\sum_{i=1}^{3}x_i(2x_4-x_i)-3x_{4}^2\right)k^2+i\left[\sum_{i=1}^{3}x_i\left(2x_{4}^2+\sum_{j=1 \atop j\neq i}^{3}x_{j}^2\right)-\right.\\
\left.4x_4\left(x_1\sum_{i=1 \atop i\neq 1}^{3}x_{i}+\prod_{i=1 \atop i\neq 1}^{3}x_{i}\right)\right]k+x_4^2\left(x_1\sum_{i=1 \atop i\neq 1}^{3}x_{i}+\prod_{i=1 \atop i\neq 1}^{3}x_{i}\right)+\prod_{i=1}^{3}x_{i}\left(\sum_{i=1}^{3}x_i-6x_4\right).
\end{multline*}
Inserting the roots of $X^2=0$ into \eqref{emme} it can be verified that $M(k)=0$ for all real $k$. Hence, we have reduced \eqref{stellastellang_x} to an Heun equation with terms $\alpha\beta$, and $q_{A}$ given by \eqref{alphabetaq}.

\section{\label{sec:5}The metric $\mathscr{B}^{0}_{+}$ : $b=c=1$,  $a=k=\gamma=0$}
In the present case we have
\begin{eqnarray*}
\epsilon_{1}&=&1,\quad\epsilon_{2}=0,\quad m(x)=-2\ell x,\quad p(w)=w^2+\ell^2,\quad T(w,x)=1,\\
fW^{2}(w)&=&g_{4}w^4+f_{2}w^2+f_{1}w+f_{0}=g_{4}\prod_{i=1}^{4}(w-w_i),\\
X^{2}(x)&=&\kappa_{2}x^2+g_{1}x+g_{0}=\kappa_{2}(x-x_1)(x-x_2),\quad \kappa_{2}=-(f_{2}+2\ell^2\Lambda)
\end{eqnarray*}
where for $i=1,\cdots,4$ $w_{i}$, and $x_1$, $x_2$ denote the roots of the polynomial equations $fW^{2}(w)=0$, and $X^{2}(x)=0$, respectively. Throughout this section we shall assume that $w_{i}\neq w_{j}$ for every $i\neq j$, and $x_1\neq x_2$. Finally, since the term $w^3$ is not present in the expression for $fW^2$ it can be easily verified that the roots of the equation $fW^2(w)=0$ satisfy the condition $w_1+w_2+w_3+w_4=0$. Taking into account the expression for $\Psi_2$ given by (2.9g) in Kamran, and McLenaghan (1984), the functions $f_s(w)$ and $g_{s}(x)$ entering, respectively in \eqref{radial}, and \eqref{angular} are computed to be 
\begin{eqnarray*}
&&f_{s}(w)=-2ir\epsilon(s)(1-2|s|)(w+i\ell)-(1-|s|)(1-2|s|)\left[\frac{{fW^2}^{'}}{w-i\ell}-\frac{6fW^2-\widehat{C}_{\Lambda}(w+i\ell)}{3(w-i\ell)^2}\right]\\
&&g_{s}(x)=0
\end{eqnarray*}
with $\widehat{C}_{\Lambda}:=-2i\ell(3g_2+4\Lambda\ell^2)+3f_1$. Concerning the ODE \eqref{radial} for $G_{s}(w)$ we find 
\begin{equation}\label{unouno_www}
\frac{d}{dw}\left(fW^2\frac{dG_{s}}{dw}\right)+\Gamma_{\ell}(w)\frac{dG_{s}}{dw}+Q_{s}(w)G_{s}=0
\end{equation}
with
\[
\Gamma_{\ell}(w)=2ir\frac{f^2-1}{f^2+1}(w^2+\ell^2),
\]
\begin{multline*}
Q_{s}(w)=\lambda_{s}-2\epsilon(s)(1-2|s|)r\ell+2ir\epsilon(s)(1-2|s|)w-\frac{2ir}{1+f^2}\left[1+\epsilon(s)-f^2(1-\epsilon(s))\right]w+\\
f|s|\left(WW^{''}+(1-|s|){W^{'}}^2\right)+2irs(w^2+\ell^2)\frac{W^{'}}{W}+\frac{4fr^2}{(1+f^2)^2}\frac{(w^2+\ell^2)^2}{W^2}+\\
(1-|s|)(1-2|s|)\left[\frac{{fW^2}^{'}}{w-i\ell}-\frac{6fW^2-\widehat{C}_{\Lambda}(w+i\ell)}{3(w-i\ell)^2}\right].
\end{multline*}
By means of the transformation $G_{s}(w)=e^{h(w)}\varphi_{s}(w)$ with the requirement that $h^{'}=-\Gamma_{\ell}/(2fW^2)$ equation \eqref{unouno_www} simplifies to
\begin{equation}\label{ultult}
\frac{d}{dw}\left(fW^2\frac{d\varphi_{s}}{dw}\right)+R_{s}(w)\varphi_{s}=0
\end{equation}
where
\begin{multline*}
R_{s}(w)=\widetilde{\lambda}_{s}-4irsw+f|s|\left(WW^{''}+(1-|s|){W^{'}}^2\right)+2irs(w^2+\ell^2)\frac{W^{'}}{W}+\frac{r^2(w^2+\ell^2)^2}{fW^2}+\\
(1-|s|)(1-2|s|)\left[\frac{{fW^2}^{'}}{w-i\ell}-\frac{6fW^2-\widehat{C}_{\Lambda}(w+i\ell)}{3(w-i\ell)^2}\right]
\end{multline*}
with
\[
\widetilde{\lambda}_{s}:=\lambda_{s}-2\epsilon(s)(1-2|s|)r\ell.
\]
For notational purposes let us introduce the function $\sigma_{\pm}(w):=w\pm \ell^2$. By means of the homographic substitution \eqref{homog} equation \eqref{ultult} becomes
\begin{equation}\label{stellang_x_www}
\frac{d^2\varphi_{s}}{dz^2}+P_{s}(z)\frac{d\varphi_{s}}{dz}+\widetilde{Q}_{s}(z)\varphi_{s}=0
\end{equation}
with
\[
P_{s}(z)=\frac{1}{z}+\frac{1}{z-1}+\frac{1}{z-z_{S}}-\frac{2}{z-z_{\infty}},
\]
\begin{multline*}
\widetilde{Q}_{s}(z)=\frac{B_{1}}{z^2}+\frac{B_{2}}{(z-1)^2}+\frac{B_{3}}{(z-z_{S})^2}+\frac{2}{(z-z_{\infty})^2}+\frac{B_{\ell}}{(z-z_{\ell})^2}+\\
\frac{A_{1}}{z}+\frac{A_{2}}{z-1}+\frac{A_{3}}{z-z_{S}}+\frac{A_{\infty}}{z-z_{\infty}}+\frac{A_{\ell}}{z-z_{\ell}},\quad z_{\ell}:=z_{\infty}\frac{w_{1}-i\ell}{w_{4}-i\ell}
\end{multline*}
where for $i=1,2,3$
\begin{eqnarray*}
&&A_{\infty}=\frac{1}{z_{\infty}(w_{4}-w_{1})}\left(\sum_{i=1}^{3}w_{i}-3w_{4}\right),\\
&&B_{i}=-\left(\frac{s}{2}-i\frac{rc_{i}~\sigma_{+}(w_{i}^2)}{w_{i}-w_{4}}\right)^2,\quad B_{\ell}=-2(1-|s|)(1-2|s|)\left(1-i\frac{\ell\widehat{C}_{\Lambda}}{3g_{4}}\prod_{j=1}^{4}(w_j-i\ell)^{-1}\right),\\
&&A_{1}=-\frac{c_{1}}{z_\infty}~\mathfrak{g}_s(w_{1}),\quad A_{2}=c_{2}\frac{w_{2}-w_{1}}{w_{4}-w_{1}}~\mathfrak{g}_s(w_{2}),\quad A_{3}=\frac{c_{3}(w_{3}-w_{1})}{z_{S}(w_{4}-w_{1})}~\mathfrak{g}_s(w_{3}),\\
&&A_{\ell}=\frac{(1-|s|)(1-2|s|)(w_4-i\ell)}{z_{\infty}(w_1-w_4)}\left(\widehat{\tau}(\ell)-i\frac{\widehat{C}_{\Lambda}}{3g_{4}}\widehat{p}(\ell)\prod_{j=1}^{4}(w_{j}-i\ell)^{-1}\right)\prod_{i=1}^{3}(w_{i}-i\ell)^{-1},
\end{eqnarray*}
with
\begin{multline*}
\mathfrak{g}_s(w_{i})=-\widetilde{\lambda}_{s}+\frac{g_4}{2}|s|p(w_{i})-\frac{2irs}{w_{i}-w_{4}}\sigma_{+}(w_{i}w_4)-\frac{2g_{4}c_{i}^{2}r^2}{w_{i}-w_{4}}\sigma_{+}(w_{i}^2)\cdot\\
\cdot\left[\sigma_{-}(w_{i}^2)\sum_{j=1 \atop j\neq i}^{3}w_{j}-2w_{i}\sigma_{-}\left(\prod_{j=1 \atop j\neq i}^{3}w_{j}~\right)\right]-\frac{(1-|s|)(1-2|s|)}{w_{i}-i\ell}\left(\frac{w_i-w_4}{c_{i}}-\frac{i}{3}\widehat{C}_{\Lambda}\frac{w_i+i\ell}{w_i-i\ell}\right),
\end{multline*}
\begin{multline*}
\widehat{\tau}(\ell):=\left(3w_{4}-\sum_{i=1}^{3}w_{i}\right)\ell^2-2i(w_{1}w_{2}+w_{1}w_{3}-w_{2}w_{4}-w_{3}w_{4}-w_{1}w_{4}+w_{2}w_{3})\ell+\\
3w_{1}w_{2}w_{3}-w_{2}w_{3}w_{4}-w_{1}w_{2}w_{4}-w_{1}w_{3}w_{4},
\end{multline*}
\begin{multline*}
\widehat{p}(\ell):=3\ell^4+i\left(\sum_{i=1}^{3}w_{i}+5w_4\right)\ell^3+\left(w_{1}w_{2}+w_{2}w_{3}+w_{1}w_{3}-3w_{4}\sum_{i=1}^{3}w_{i}\right)\ell^2-\\
i(w_1w_2w_4+w_1w_3w_4+w_2w_3w_4-3w_1w_2w_3)\ell-w_1w_2w_3w_4,
\end{multline*}
constants $c_{i}$ defined as in Section~\ref{sec:3}, and $p(w_{i})$ given by \eqref{polp}. Inserting the roots of the polynomial equation $fW^2(w)=0$ into the expressions for $B_\ell$, and $A_\ell$ we checked with the help of Maple 9.5 that $B_\ell=A_\ell=0$. In order to reduce \eqref{stellang_x_www} to a Heun equation we just need to introduce the F-homotopic transformation
\[
\varphi_{s}(z)=z^{\alpha_{1}}(z-1)^{\alpha_{2}}(z-z_{S})^{\alpha_{3}}(z-z_{\infty})\widetilde{\varphi}_{s}(z).
\]
with exponents $\alpha_{i}$ such that ${\alpha_{i}}^{2}=-B_{i}$, and to proceed as we did for the equation \eqref{stellastellang_x} in Section~\ref{sec:4}. Concerning the equation for $H_{s}(x)$ we find
\begin{multline}\label{unoB0md}
\frac{d}{dx}\left(X^2\frac{dH_{s}}{dx}\right)+\left[-\lambda_{s}+2\epsilon(s)\ell r-2s(2\ell rx+q)\frac{X^{'}}{X}+\right.\\
\left. |s|\left(XX^{''}+(1-|s|){X^{'}}^{2}\right)-\frac{(2\ell rx+q)^2}{X^2}\right]H_{s}=0.
\end{multline}
By means of the transformation $z=(x-x_{1})/(x_{2}-x_{1})$ mapping the points $x_{1},x_{2},\infty$ to $0,1,\infty$ our equation \eqref{unoB0md} becomes
\begin{equation}\label{stellaB0md}
\frac{d^2H_{s}}{dz^2}+\mathfrak{P}_{s}(z)\frac{dH_{s}}{dz}+\mathfrak{Q}_{s}(z)H_{s}=0
\end{equation}
with
\[
\mathfrak{P}_{s}(z)=\frac{1}{z}+\frac{1}{z-1},\quad
\mathfrak{Q}_{s}(z)=\frac{\widetilde{B}_{1}}{z^2}+\frac{\widetilde{B}_{2}}{(z-1)^2}+\frac{\widetilde{A}_{1}}{z}+\frac{\widetilde{A}_{2}}{z-1}
\]
where
\begin{eqnarray*}
&&\widetilde{B}_{1}=-\left(\frac{s}{2}-\frac{2r\ell x_{1}+q}{\kappa_{2}(x_{2}-x_{1})}\right)^2,\quad\widetilde{B}_{2}=-\left(\frac{s}{2}+\frac{2r\ell x_{2}+q}{\kappa_{2}(x_{2}-x_{1})}\right)^2,\\
&&\widetilde{A}_{1}=-\frac{\lambda_{s}-2\ell r(\epsilon(s)-s)}{\kappa_{2}}+\frac{|s|(|s|-2)}{2}-\frac{2}{\kappa_{2}^2(x_2-x_1)^2}\prod_{i=1}^{2}(2r\ell x_{i}+q),\quad \widetilde{A}_{2}=-\widetilde{A}_{1}.
\end{eqnarray*}
If we make the F-homotopic transformation
\[
H_{s}(z)=z^{\widehat{\alpha}_{1}}(z-1)^{\widehat{\alpha}_{2}}\widetilde{H}_{s}(z)
\]
with the requirement that ${\widehat{\alpha}_{i}}^{2}=-\widetilde{B}_{i}$ then \eqref{stellaB0md} reads
\begin{equation}\label{stellastellaB0md}
\widetilde{H}_{s}^{''}(z)+\widehat{\mathfrak{P}}_{s}(z)\widetilde{H}_{s}^{'}(z)+\widehat{\mathfrak{Q}}_{s}(z)\widetilde{H}_{s}(z)=0
\end{equation}
with
\[
\widehat{\mathfrak{P}}_{s}(z)=\frac{1+2\widehat{\alpha}_{1}}{z}+\frac{1+2\widehat{\alpha}_{2}}{z-1},\quad
\widehat{\mathfrak{Q}}_{s}(z)=\frac{(\widetilde{A}_{1}+\widetilde{A}_{2})z+\widehat{\alpha}_{2}+\widehat{\alpha}_{1}(1+2\widehat{\alpha}_{2})-\widetilde{A}_{1}}{z(z-1)}.
\]
Since $\widetilde{A}_{2}=-\widetilde{A}_{1}$ it follows that \eqref{stellastellaB0md} is an HYE.

\section{\label{sec:6}The metric $\mathscr{C}^{*}$ : $a=1$,  $c=\sqrt{2}$,  $b=k=\ell=0$ , $\gamma=\pi/4$}
In this case we have
\begin{eqnarray*}
\epsilon_{1}&=&0,\quad\epsilon_{2}=\frac{\sqrt{2}}{2},\quad m(x)=-\sqrt{2}~x^2,\quad p(w)=0,\quad T(w,x)=wx+1,\\
fW^{2}(w)&=&\sum_{n=0}^{3}f_{n}w^{n}=f_{3}\prod_{i=1}^{3}(w-w_{i}),\quad X^{2}(x)=g_{4}x^4+f_{1}x^3-f_{2}x^2+g_{1}x+g_{0}.
\end{eqnarray*}
Taking into account that \eqref{parametri0}, and \eqref{parametri1} imply in the present framework that
\[
g_{1}=f_{3},\quad g_{0}=0,
\]
the function $X^{2}(x)$ can be written as follows
\[
X^{2}(x)=x\left(g_{4}x^3+f_{1}x^{2}-f_{2}x+f_{3}\right)=g_{4}x\prod_{i=1}^{3}(x-x_{i}).
\]
Moreover, for $i$,$j=1,\cdots,3$ let $w_{i}$ and $x_{i}$ denote the i-th root of the polynomial equations $fW^{2}(w)=0$, and $X^{2}(x)=0$, respectively. Throughout this section we shall assume that $w_{i}\neq w_{j}$, and $x_{i}\neq x_{j}$ for every $i\neq j$. Furthermore, in the present case \eqref{cosmological} reads
\[
g_{4}=-\left(\frac{\Lambda}{3}+f_{0}\right).
\]
Finally, the roots of $X^{2}(x)=0$ satisfy the following useful relations
\[
\sum_{i=1}^{3}x_{i}=-\frac{f_1}{g_4},\quad \prod_{i=1}^{3}x_{i}=-\frac{f_3}{g_4}.
\]
Making use of the expression for $\Psi_{2}$ given by (2.10g) in Kamran, and McLenaghan (1984), the functions $f_{s}(w)$, and $g_{s}(x)$ entering, respectively in \eqref{radial}, and \eqref{angular} are computed to be
\begin{eqnarray*}
&&f_{s}(w)=-f_{3}(1-|s|)(1-2|s|)w,\\
&&g_{s}(x)=2\sqrt{2}r\epsilon(s)(1-2|s|)x-\frac{(1-|s|)(1-2|s|)}{x}\left[\frac{dX^2}{dx}-\frac{2}{x}X^2+f_{3}\right].
\end{eqnarray*}
Hence, the equation for $G_{s}(w)$ becomes
\begin{equation}\label{uno_s_1}
\frac{d}{dw}\left(fW^2\frac{dG_{s}}{dw}\right)+\Gamma_{q}\frac{dG_{s}}{dw}+Q_{s}(w)G_{s}=0
\end{equation}
with
\[
\Gamma_{q}:=i\sqrt{2}q\frac{1-f^2}{1+f^2},
\]
\begin{multline*}
Q_{s}(w):=\lambda_{s}+f_{3}(1-|s|)(1-2|s|)w+f|s|\left(WW^{''}+(1-|s|){W^{'}}^2\right)-\\
i\sqrt{2}sq\frac{W^{'}}{W}+\frac{2fq^2}{(1+f^2)^2}\frac{1}{W^2}.
\end{multline*}
Equation \eqref{uno_s_1} can be further simplified. To this aim let us transform $G_{s}(w)$ according to
\[
G_{s}(w)=e^{h(w)}\varphi_{s}(w).
\] 
If we require that $h^{'}=-\Gamma_{q}/(2fW^2)$ we obtain the following equation for $\varphi_{s}(w)$
\begin{equation}\label{due_s_1}
\frac{d}{dw}\left(fW^2\frac{d\varphi_{s}}{dw}\right)+R_{s}(w)~\varphi_{s}=0
\end{equation}
where
\begin{multline*}
R_{s}(w):=\lambda_{s}+f_{3}(1-|s|)(1-2|s|)w+f|s|\left(WW^{''}+(1-|s|){W^{'}}^2\right)-\\
i\sqrt{2}sq\frac{W^{'}}{W}+\frac{q^2}{2fW^2}.
\end{multline*}
For notational purposes let us define constants
\[
c_{i}^{-1}:=f_{3}\prod_{j=1 \atop j\neq i}^{3}(w_{i}-w_{j}),\quad i=1,2,3.
\]
By means of the homographic substitution
\begin{equation}\label{homog_1}
z=\frac{w-w_{1}}{w_2-w_{1}}
\end{equation}
mapping the points $w_{1},w_{2},w_{3},\infty$ to $0,1,z_{S},\infty$ with $z_{S}:=(w_{3}-w_{1})/(w_{2}-w_{1})$, equation \eqref{due_s_1} becomes
\begin{equation}\label{stella_s_11}
\frac{d^2\varphi_{s}}{dz^2}+P(z)\frac{d\varphi_{s}}{dz}+\widetilde{Q}_{s}(z)\varphi_{s}=0
\end{equation}
with
\begin{eqnarray*}
&&P(z)=\frac{1}{z}+\frac{1}{z-1}+\frac{1}{z-z_{S}},\\
&&\widetilde{Q}_{s}(z)=\frac{B_{1}}{z^2}+\frac{B_{2}}{(z-1)^2}+\frac{B_{3}}{(z-z_{S})^2}+\frac{A_{1}}{z}+\frac{A_{2}}{z-1}+\frac{A_{3}}{z-z_{S}}
\end{eqnarray*}
where for $i=1,2,3$
\begin{eqnarray*}
&&B_{i}=-\left(\frac{s}{2}+i\frac{qc_{i}}{\sqrt{2}}\right)^2,\quad A_{i}=-(w_1-w_2)c_{i}\mathfrak{g}_{s}(w_{i}),\\
&&\mathfrak{g}_{s}(w_{i})=\lambda_{s}+f_{3}\left[(1-|s|)(1-2|s|)w_{i}-\left(\frac{|s|}{2}(2-|s|)-q^2c_{i}^2\right)\left(\sum_{j=1 \atop j\neq i}^{3}w_{j}-2w_{i}\right)\right].
\end{eqnarray*}
If we make the F-homotopic transformation
\[
\varphi_{s}(z)=z^{\alpha_{1}}(z-1)^{\alpha_{2}}(z-z_{S})^{\alpha_{3}}\widetilde{\varphi}_{s}(z)
\]
and require that $\alpha_{i}^{2}=-B_{i}$, then \eqref{stella_s_11} becomes
\begin{equation}\label{stellastella_stella}
\frac{d^2\widetilde{\varphi}_{s}}{dz^2}+\widehat{P}_{s}(z)\frac{d\widetilde{\varphi}_{s}}{dz}+\widehat{Q}_{s}(z)\widetilde{\varphi}_{s}=0
\end{equation}
with
\[
\widehat{P}_{s}(z)=\frac{1+2\alpha_{1}}{z}+\frac{1+2\alpha_{2}}{z-1}+\frac{1+2\alpha_{3}}{z-z_{S}},\quad
\widehat{Q}_{s}(z)=\frac{\widehat{A}_{1}}{z}+\frac{\widehat{A}_{2}}{z-1}+\frac{\widehat{A}_{3}}{z-z_{S}}.
\]
Finally, it can be checked that $\widehat{A}_{1}+\widehat{A}_{2}+\widehat{A}_{3}=0$ with
\begin{eqnarray*}
\widehat{A}_{1}&=&A_{1}-\alpha_{1}-\alpha_{2}(1+2\alpha_{1})-\frac{\alpha_{1}+\alpha_{3}(1+2\alpha_{1})}{z_{S}},\\
\widehat{A}_{2}&=&A_{2}+\alpha_{1}+\alpha_{2}(1+2\alpha_{1})-\frac{\alpha_{2}+\alpha_{3}(1+2\alpha_{2})}{z_{S}-1},\\
\widehat{A}_{3}&=&A_{3}+\frac{\alpha_{2}+\alpha_{3}(1+2\alpha_{2})}{z_{S}-1}+\frac{\alpha_{1}+\alpha_{3}(1+2\alpha_{1})}{z_{S}}.
\end{eqnarray*}
Hence, \eqref{stellastella_stella} reduces to an Heun equation with terms $\alpha\beta$ and accessory parameter $q_{A}$ given by \eqref{parametri}. Concerning \eqref{angular} we find the following equation for $H_{s}(x)$, namely
\begin{multline}\label{unouno_s20}
\frac{d}{dx}\left(X^2\frac{dH_{s}}{dx}\right)+\left[-\lambda_{s}+4\sqrt{2}rsx+|s|\left(XX^{''}-(|s|-1){X^{'}}^2\right)-2\sqrt{2}rsx^2\frac{X^{'}}{X}-2r^2\frac{x^4}{X^2}+\right.\\
\left.(1-|s|)(1-2|s|)\left(\frac{{X^{2}}^{'}}{x}-2\frac{X^{2}}{x^2}+\frac{f_3}{x}\right)\right]H_{s}=0.
\end{multline}
Let us introduce the following notation
\[
\widehat{c}_{1}~^{-1}=g_{4}x_1x_2,\quad\widehat{c}_{2}~^{-1}=g_4x_1(x_1-x_2),\quad\widehat{c}_{3}~^{-1}=g_4x_2(x_2-x_1) .
\]
By means of the homographic transformation
\begin{equation}\label{homog2_s20}
z=\frac{x}{x-x_{3}}~\frac{x_{1}-x_{3}}{x_{1}}
\end{equation}
mapping the points $0,x_{1},x_{2},x_{3},\infty$ to $0,1,x_{S},\infty,x_{\infty}$ with
\begin{equation}\label{xI_xs_s20}
x_{\infty}=\frac{x_{1}-x_{3}}{x_{1}},\quad x_{S}=\frac{x_{2}x_{\infty}}{x_{2}-x_{3}}
\end{equation}
our equation \eqref{unouno_s20} becomes
\begin{equation}\label{stellang}
\frac{d^2H_{s}}{dz^2}+\mathfrak{P}_{s}(z)\frac{dH_{s}}{dz}+\mathfrak{Q}_{s}(z)H_{s}=0
\end{equation}
with
\begin{eqnarray*}
&&\mathfrak{P}_{s}(z)=\frac{1}{z}+\frac{1}{z-1}+\frac{1}{z-x_{S}}-\frac{2}{z-x_{\infty}},\\
&&\mathfrak{Q}_{s}(z)=\frac{\widetilde{B}_{1}}{z^2}+\frac{\widetilde{B}_{2}}{(z-1)^2}+\frac{\widetilde{B}_{3}}{(z-x_{S})^2}+\frac{2}{(z-x_{\infty})^2}+\frac{\widetilde{A}_{1}}{z}+\frac{\widetilde{A}_{2}}{z-1}+\frac{\widetilde{A}_{3}}{z-x_{S}}+\frac{\widetilde{A}_{\infty}}{z-x_{\infty}}
\end{eqnarray*}
where
\begin{eqnarray*}
&&\widetilde{A}_{\infty}=\frac{x_1+x_2-3x_3}{x_3x_{\infty}},\\
&&\widetilde{B}_{1}=-\frac{s^2}{4},\quad\widetilde{B}_{2}=-\left(\frac{s}{2}+\sqrt{2}r\frac{\widehat{c}_{2}x_{1}^2}{x_{1}-x_{3}}\right)^2,\quad\widetilde{B}_{3}=-\left(\frac{s}{2}+\sqrt{2}r\frac{\widehat{c}_{3}x_{2}^2}{x_{2}-x_{3}}\right)^2\\
&&\widetilde{A}_{1}=-\frac{\widehat{c}_{1}}{x_{\infty}}~\widehat{\mathfrak{g}}_{s}^{I},\quad \widetilde{A}_{2}=\frac{\widehat{c}_{2}x_{1}}{x_{3}}~\widehat{\mathfrak{g}}_{s}^{II},\quad \widetilde{A}_{3}=\frac{\widehat{c}_{3}x_{2}}{x_{3}x_{S}}~\widehat{\mathfrak{g}}_{s}^{III},\quad
\widehat{\mathfrak{g}}_{s}^{I}=\lambda_{s}+\frac{g_4}{2}|s|A,
\end{eqnarray*}
\begin{multline*}
\widehat{\mathfrak{g}}_{s}^{II}=\lambda_{s}+\frac{g_{4}}{2}|s|B+2\sqrt{2}rs\frac{x_1x_3}{x_1-x_3}-(1-|s|)(1-2|s|)\left[g_4(x_1-x_3)(x_1-x_2)+f_3x_1\right]-\\
4r^2~\widehat{c}_2^{2}g_4\frac{x_1^4x_2}{x_1-x_3},
\end{multline*}
\begin{multline*}
\widehat{\mathfrak{g}}_{s}^{III}=\lambda_{s}+\frac{g_{4}}{2}|s|C+2\sqrt{2}rs\frac{x_2x_3}{x_2-x_3}-(1-|s|)(1-2|s|)\left[g_4(x_2-x_3)(x_2-x_1)-\frac{f_3}{x_2}\right]-\\
4r^2~\widehat{c}_3^{2}g_4\frac{x_1x_2^4}{x_2-x_3},
\end{multline*}
\begin{eqnarray*}
&&A:=(|s|-2)x_3(x_1+x_2)+2(|s|-1)x_1x_2,\\
&&B:=2(2|s|-3)x_1^2+(4-3|s|)x_1x_2+(|s|-2)x_3(x_2-2x_1),\\
&&C:=2(2|s|-3)x_2^2+(4-3|s|)x_1x_2+(|s|-2)x_3(x_1-2x_2 ).
\end{eqnarray*}
If we make the F-homotopic transformation
\[
H_{s}(z)=z^{\widehat{\alpha}_{1}}(z-1)^{\widehat{\alpha}_{2}}(z-x_{S})^{\widehat{\alpha}_{3}}(z-x_{\infty})\widetilde{H}_{s}(z)
\]
together with the requirement that $\widehat{\alpha}_{i}^{2}=-\widetilde{B}_{i}$, then \eqref{stellang} becomes
\begin{equation}\label{stellastellang_00}
\frac{d^2\widetilde{H}_{s}}{dz^2}+\widehat{\mathfrak{P}}_{s}(z)\frac{d\widetilde{H}_{s}}{dz}+\widehat{\mathfrak{Q}}_{s}(z)\widetilde{H}_{s}=0
\end{equation}
with
\[
\widehat{\mathfrak{P}}_{s}(z)=\frac{1+2\widehat{\alpha}_{1}}{z}+\frac{1+2\widehat{\alpha}_{2}}{z-1}+\frac{1+2\widehat{\alpha}_{3}}{z-x_{S}},\quad
\widehat{\mathfrak{Q}}_{s}(z)=\frac{\mathfrak{A}_{1}}{z}+\frac{\mathfrak{A}_{2}}{z-1}+\frac{\mathfrak{A}_{3}}{z-x_{S}}.
\]
Now, employing the expressions for the coefficients $\widetilde{A}_{i}$ with $i=1,2,3$ it can be checked that
\[
\sum_{i=1}^{3}\mathfrak{A}_{i}=\sum_{i=1}^{3}\widetilde{A}_{i}-\left(\frac{1}{x_{\infty}}+\frac{1}{x_{\infty}-x_{S}}+\frac{1}{x_{\infty}-1}\right)=0.
\]
Hence, \eqref{stellastellang_00} becomes an Heun equation with
\[
\alpha\beta=x_{S}\mathfrak{A}_{3}+\mathfrak{A}_{2},\quad q_{A}=-x_{S}\mathfrak{A}_{1}
\]
where again the $\mathfrak{A}_{i}$'s can be directly obtained from the formulae \eqref{I}-\eqref{III} for the coefficients $\widehat{A}_{i}$ by means of the formal substitutions $\widehat{A}_{i}\longrightarrow\mathfrak{A}_{i}$, $A_{i}\longrightarrow \widetilde{A}_{i}$, $\alpha_{i}\longrightarrow \widehat{\alpha}_{i}$, $z_{S}\longrightarrow x_{S}$ and $z_{\infty}\longrightarrow x_{\infty}$.

\section{\label{sec:7}The metric $\mathscr{C}^{00}$ : $b=\ell=1$,  $a=c=k=\gamma=0$ }
In this case we have
\begin{eqnarray*}
\epsilon_{1}&=&1=p(w)=T(w,x),\quad\epsilon_{2}=0=m(x),\\
fW^{2}(w)&=&\sum_{n=0}^{2}f_{n}w^{n}=f_{2}(w-w_{1})(w-w_2),\\
X^{2}(x)&=&-\Lambda x^2+g_{1}x+g_{0}=-\Lambda(x-x_{1})(x-x_2).
\end{eqnarray*}
Throughout this section we shall assume that $w_{1}\neq w_{2}$, and $x_{1}\neq x_{2}$. Making use of the expression for $\Psi_{2}$ given by (2.11g) in Kamran, and McLenaghan (1984), the functions $f_{s}(w)$, and $g_{s}(x)$ entering, respectively in \eqref{radial}, and \eqref{angular} are computed to be
\[
f_{s}(w)=\frac{2}{3}\Lambda(1-|s|)(1-2|s|),\quad g_{s}(x)=0.
\]
The equation for $G_{s}(w)$ becomes
\begin{equation}\label{uno_s_pp}
\frac{d}{dw}\left(fW^2\frac{dG_{s}}{dw}\right)+\widehat{\Gamma}\frac{dG_{s}}{dw}+Q_{s}(w)G_{s}=0
\end{equation}
with $\widehat{\Gamma}:=2ir(f^2-1)/(f^2+1)$, and
\begin{multline*}
Q_{s}(w):=\lambda_{s}-\frac{2}{3}\Lambda (1-|s|)(1-2|s|)+2irs\frac{W^{'}}{W}+\\
f|s|\left(WW^{''}+(1-|s|){(W^{'})}^2\right)+\frac{4r^2 f}{(1+f^2)^2}\frac{1}{W^2}.
\end{multline*}
By means of the transformation $G_{s}(w)=e^{h(w)}\varphi_{s}(w)$ with the requirement that $h^{'}=-\Gamma/(2fW^2)$ we obtain the following equation for $\varphi_{s}(w)$
\begin{equation}\label{due_s_pp}
\frac{d}{dw}\left(fW^2\frac{d\varphi_{s}}{dw}\right)+\widehat{Q}_{s}(w)~\varphi_{s}=0
\end{equation}
where
\[
\widehat{Q}_{s}(w):=\lambda_{s}-\frac{2}{3}\Lambda (1-|s|)(1-2|s|)+2irs\frac{W^{'}}{W}+
f|s|\left(WW^{''}+(1-|s|){(W^{'})}^2\right)+\frac{r^2}{fW^2}.
\]
By means of the variable transformation $z=(w-w_{1})/(w_2-w_{1})$ mapping the points $w_{1},w_{2},\infty$ to $0,1,\infty$ equation \eqref{due_s_pp} becomes
\begin{equation}\label{stella_pp}
\frac{d^2\varphi_{s}}{dz^2}+P(z)\frac{d\varphi_{s}}{dz}+\widetilde{Q}_{s}(z)\varphi_{s}=0
\end{equation}
with
\[
P(z)=\frac{1}{z}+\frac{1}{z-1},\quad \widetilde{Q}_{s}(z)=\frac{B_{1}}{z^2}+\frac{B_{2}}{(z-1)^2}+\frac{A_{1}}{z}+\frac{A_{2}}{z-1}
\]
where
\begin{eqnarray*}
&&B_{1}=-\left(\frac{s}{2}+i\frac{r}{f_{2}(w_{2}-w_{1})}\right)^2,\quad B_{2}=-\left(\frac{s}{2}-i\frac{r}{f_{2}(w_{2}-w_{1})}\right)^2,\\
&&A_{1}=-\frac{\lambda_{s}-(2/3)\Lambda(1-|s|)(1-2|s|)}{f_{2}}+\frac{|s|(|s|-2)}{2}+\frac{2r^2}{f_{2}^{2}(w_2-w_1)^2},\quad A_{2}=-A_{1}.
\end{eqnarray*}
If we make the F-homotopic transformation
\[
\varphi_{s}(z)=z^{\alpha_{1}}(z-1)^{\alpha_{2}}\widetilde{\varphi}_{s}(z)
\]
and require that $\alpha_{i}^{2}=-B_{i}$ for $i=1,2$, then \eqref{stella_pp} becomes
\begin{equation}\label{stellastella_pp}
\frac{d^2\widetilde{\varphi}_{s}}{dz^2}+\widehat{P}_{s}(z)\frac{d\widetilde{\varphi}_{s}}{dz}+\widehat{Q}_{s}(z)\varphi_{s}=0
\end{equation}
with
\[
\widehat{P}_{s}(z)=\frac{1+2\alpha_{1}}{z}+\frac{1+2\alpha_{2}}{z-1},\quad
\widehat{Q}_{s}(z)=\frac{(A_1+A_2)z+\alpha_{2}+\alpha_{1}(1+2\alpha_{2})-A_{1}}{z(z-1)}.
\]
Since $A_{2}=-A_{1}$ it follows that \eqref{stellastella_pp} is an HYE. Concerning \eqref{angular} we find the following equation for $H_{s}(x)$, namely
\begin{equation}\label{unouno_pp}
\frac{d}{dx}\left(X^2\frac{dH_{s}}{dx}\right)+\left[-\lambda_{s}+|s|\left(XX^{''}+(1-|s|){X^{'}}^2\right)-2sq\frac{X^{'}}{X}-\frac{q^2}{X^2}\right]H_{s}=0.
\end{equation}
By means of the transformation $z=(x-x_{1})(x_{2}-x_{1})$ mapping the points $x_{1},x_{2},\infty$ to $0,1,\infty$ our equation \eqref{unouno_pp} becomes
\begin{equation}\label{stellang_pp}
\frac{d^2H_{s}}{dz^2}+\mathfrak{P}_{s}(z)\frac{dH_{s}}{dz}+\mathfrak{Q}_{s}(z)H_{s}=0
\end{equation}
with
\[
\mathfrak{P}_{s}(z)=\frac{1}{z}+\frac{1}{z-1},\quad \mathfrak{Q}_{s}(z)=\frac{\widetilde{B}_{1}}{z^2}+\frac{\widetilde{B}_{2}}{(z-1)^2}+\frac{\widetilde{A}_{1}}{z}+\frac{\widetilde{A}_{2}}{z-1}
\]
where
\begin{eqnarray*}
&&\widetilde{B}_{1}=-\left(\frac{s}{2}+\frac{q}{\Lambda(x_{2}-x_{1})}\right)^2,\quad\widetilde{B}_{2}=-\left(\frac{s}{2}-\frac{q}{\Lambda(x_{2}-x_{1})}\right)^2,\\
&&\widetilde{A}_{1}=-\frac{\lambda_{s}}{\Lambda}+\frac{|s|(|s|-2)}{2}-\frac{2q^2}{\Lambda^2(x_2-x_1)^2},\quad \widetilde{A}_{2}=-\widetilde{A}_{1}. 
\end{eqnarray*}
Finally, with the F-homotopic transformation
\[
H_{s}(z)=z^{\widehat{\alpha}_{1}}(z-1)^{\widehat{\alpha}_{2}}\widetilde{H}_{s}(z),
\]
and the requirement that $\widehat{\alpha}_{i}^{2}=-\widetilde{B}_{i}$ for $i=1,2$ \eqref{stellang_pp} becomes
\begin{equation}\label{stellastellang_pp}
\frac{d^2\widetilde{H}_{s}}{dz^2}+\widehat{\mathfrak{P}}_{s}(z)\frac{d\widetilde{H}_{s}}{dz}+\widehat{\mathfrak{Q}}_{s}(z)\widetilde{H}_{s}=0
\end{equation}
with
\[
\widehat{\mathfrak{P}}_{s}(z)=\frac{1+2\widehat{\alpha}_{1}}{z}+\frac{1+2\widehat{\alpha}_{2}}{z-1},\quad
\widehat{\mathfrak{Q}}_{s}(z)=\frac{(\widetilde{A}_1+\widetilde{A}_2)z+\widehat{\alpha}_{2}+\widehat{\alpha}_{1}(1+2\widehat{\alpha}_{2})-\widetilde{A}_{1}}{z(z-1)}
\]
Taking into account that $\widetilde{A}_{2}=-\widetilde{A}_{1}$ it results that \eqref{stellastellang_pp} is an HYE.

\begin{acknowledgments}
The research of D.B. was supported by the EU grant HPRN-CT-2002-00277. One of us (D.B.) is indebted to N. Kamran, McGill University, Montreal, Canada and S.Q. Wu, Hua-Zhong Normal University, Wuhan, China in finding some useful references.
\end{acknowledgments}


\begin{references}
\item[[1]\hspace{-2mm}]
Arenstorf,~R.~F., Cohen,~J.~M., and Kegeles,~L.~S. ''Electromagnetic radiation near black holes and neutron stars'' ', J. Math. Phys. \textbf{19}, 833 (1978)
\item[[2]\hspace{-2mm}]
Bardeen,~J.~M., and Press,~W.~H., ''Radiation fields in the Schwarzschild background'' ', J. Math. Phys. \textbf{14}, 7 (1973)
\item[[3]\hspace{-2mm}]
Bertotti,~B., ''Uniform Electromagnetic Field in the Theory of General Relativity'' ', Phys. Rev. \textbf{116}, 1331 (1959)
\item[[4]\hspace{-2mm}]
Blandin,~J., Pons,~R., and Marcilhacy, G., ''General Solution of Teukolsky's equation'' ', Lett. Nuovo Cimento \textbf{38}, 561 (1983)
\item[[5]\hspace{-2mm}]
Breuer,~R.~A.,~Ryan,~M.~P., and Waller,~S., ''Some properties of spin-weighted spheroidal harmonics'' ', Proc. R. Soc. London A \textbf{358}, 71 (1977)
\item[[6]\hspace{-2mm}]
Carter,~B., ''Hamilton-Jacobi and Schr\"{o}dinger separable solutions of Einstein's equations'' ', Comm. Math. Phys. \textbf{10}, 280 (1968)
\item[[7]\hspace{-2mm}]
Chakrabarti,~S.~K., ''On mass-dependent spheroidal harmonics of spin one-half'' ', Proc. R. Soc. London A \textbf{391}, 27 (1984)
\item[[8]\hspace{-2mm}]
Fackerell,~E.~D., and Crossman,~R.~G., ''Spin-weighted angular spheroidal functions'' ', J. Math. Phys. \textbf{18}, 1849 (1977)
\item[[9]\hspace{-2mm}]
Heun,~K., ''Zur Theorie der Riemmanschen Functionen Zweiter Ordnung mit vier Verzweigungspunkten'' ', Mathematische Annalen \textbf{33}, 161 (1889)
\item[[10]\hspace{-2mm}]
Kamran,~N., and McLenaghan,~R.~G., ''Exhaustive integration and a single expression for the general solution of the type D vacuum and electrovac field equations with cosmological constant for a nonsingular aligned Maxwell field'' ', J.~Math.~Phys. \textbf{25}, 1955 (1984)
\item[[11]\hspace{-2mm}]
Kamran,~N., and McLenaghan,~R.~G., ''Separation of variables for higher spin, zero rest-mass field equations on type $D$ vacuum backgrounds with cosmological constant'' 'in \textit{Gravitation and geometry}, (Monogr.~Textbooks Phys.~Sci., 4, Bibliopolis, Naples), 279  (1987)
\item[[12]\hspace{-2mm}]
Levi-Civita,~T., ''Il sottocaso $B_{2}$: soluzioni oblique'' ', Rend. Acc. Lincei \textbf{27}, 343 (1918)
\item[[13]\hspace{-2mm}]
Leahy,~D.~A., and Unruh,~W.~G., ''Angular dependence of neutrino emission from rotating black hole'' ', Phys. Rev. D \textbf{19}, 3509 (1979)
\item[[14]\hspace{-2mm}]
Leaver,~E.~W., ''Solutions to a generalized spheroidal wave equation: Teukolsky's equations in general relativity, and the two-center problem in molecular quantum mechanics'' ', J. Math. Phys. \textbf{27}, 1238 (1986)
\item[[15]\hspace{-2mm}]
Lee,~C.~H., ''Coupled gravitational and electromagnetic perturbations around a charged black hole'' ', J. Math. Phys. \textbf{17}, 1226 (1976)
\item[[16]\hspace{-2mm}]
Page,~D.~N., ''Particle emission rates from a black hole: Massless particles from an uncharged, non rotating hole'' ', Phys. Rev. D \textbf{13}, 198 (1973)
\item[[17]\hspace{-2mm}]
Plebanski,~J.~F., and Demianski,~M., ''Rotating, Charged, and Uniformly Accelerating Mass in General Relativity'' ', Ann. Phys. \textbf{98}, 98 (1976)
\item[[18]\hspace{-2mm}]
Press,~W.~H., and Teukolsky,~S.~A., ''Perturbations of a Rotating Black Hole. 2. Dynamical Stability of the Kerr metric'' ', Astrophys. J. \textbf{185}, 649 (1973)
\item[[19]\hspace{-2mm}]
Robinson,~I., ''A solution of the Maxwell-Einstein Equations'' ', Bull. Acad. Polon. Sci. \textbf{7}, 351 (1959)
\item[[20]\hspace{-2mm}]
Ronveaux,~A., \textit{Heun's Differential Equations}, (Oxford University Press, Oxford, New York, Tokyo, 1995)
\item[[21]\hspace{-2mm}]
Seidel,~E., ''A comment on the eigenvalues of spin-weighted spheroidal function'' ', Class. Quant. Grav. \textbf{6}, 1057 (1989)
\item[[22]\hspace{-2mm}]
Suzuki,~H., Takasugi,~E. and Umetsu,~H. ''Perturbations of Kerr-de Sitter Black Holes and Heun's Equations'' ', Prog. Theor. Phys. \textbf{100}, 491 (1998)
\item[[23]\hspace{-2mm}]
Wald, R.M., \textit{General Relativity}, (The University of Chicago Press, Chicago, 1984)
\item[[24]\hspace{-2mm}]
Wu,~S.~Q.,and Cai,~X., ''Massive Complex Scalar Field in a Kerr-Sen Black Hole Background: Exact Solution of Wave Equation and Hawking radiation'' ', J. Math. Phys. \textbf{44}, 1084 (2003)

\end{references}
\end{document}